# ROLE OF GENETIC POLYMORPHISMS IN TRANSGENERATIONAL INHERITANCE IN BUDDING YEAST

**Zuobin Zhu, Qing Lu，Dejian Yuan，Yanke Li，Xian Man，Yueran Zhu， and Shi Huang***

[1]State Key Laboratory of Medical Genetics, Central South University, 110 Xiangya Road, Changsha, Hunan 410078, P.R. China

*Corresponding author: email: huangshi@sklmg.edu.cn, tel 86-731-84805342

We studied the effect of single nucleotide polymorphisms (SNPs) on transgenerational inheritance of yeast segregants that were derived from a cross between a laboratory strain and a wild strain of *Saccharomyces cerevisiae*. We compared segregants with high minor allele content (MAC) relative to those with less and found a more dramatic shortening of the lag phase length for the high MAC group in response to 14 days of ethanol training. Also, the short lag phase as acquired and epigenetically memorized by ethanol training was more dramatically lost after 7 days of recovery in ethanol free medium for the high MAC group. We also found MAC linkage to mRNA expression of hundreds of genes and a preferential effect of MAC on traits with high number of known additive quantitative trait loci (QTLs). The study may help explain human variations in disease susceptibility and the missing heritability problem in complex traits/diseases.

*Abbreviations*:

MAF: minor allele frequency

MAC: minor allele content

HMAC: high minor allele content

LMAC: low minor allele content

QTL： quantitative trait loci

Different cell types of an individual organism carry the same DNA but manifest different traits or functions due to different epigenetic programing. A cell or organism may also acquire new traits by way of epigenetic reprograming through interaction with the environment. It is well established that both inherent traits and acquired traits can be transmitted through multiple generations with some traits more stable than others [1,2]. But the relationship between the stability of such transgenerational inheritance and the degree of genetic variations in an individual or cell has yet to be explored. Also unknown is the relationship between genetic variations and sensitivity to environmental factors. Better understanding of such relationships may help explain the well-known variations in disease susceptibility in human individuals when exposed to the same environmental pathogenic factors.

We here asked whether excess genetic variations can affect the transgenrational epigenetic inheritance of a trait in responses to environmental factors. We used a panel of 124 yeast segregants derived from a cross between a laboratory strain BY4716 and a vineyard isolate RM11-1a of *Saccharomyces cerevisiae*. Each segregant is homozygous in nearly all SNPs and the study here used a panel of 2956 SNPs previously genotyped for these segregants [3,4]. For a given panel of segregants, we called minor alleles (MAs) as those parental alleles that were carried by less than half of the strains in the panel. The strains would differ in the contents of MAs that each carries, and we defined “MA contents or MAC” as the total number of MAs in an individual divided by the number of SNPs scanned. Different from MA frequency (MAF), MAC is an individual measure and represents the amount of MAs in an individual of a population where MAF calculation was based to call an allele a minor one.

The genetic or SNP difference that exists between the parental BY and RM strains are due to random mutations. In a panel of segregants, some parental alleles would be less represented or found as MAs. This could be due to random drift. Alternatively, such alleles may be slightly deleterious to certain segregants and under slightly more negative selection than positive during the growth and propagation of segregants under laboratory conditions. Furthermore, if a segregant is enriched with these deleterious MAs by chance, it would be expected to have properties more likely to be under slightly negative selection. Given that a trait is typically an outcome of highly ordered biochemical processes, one predicts that segregants with more MAC should have lower capacity to maintain stable inheritance of a trait, if the reason for those MAs to be minor in the segregants panel is because they are slightly deleterious. On the other hand, if these MAs are minor simply because of arbitrary random drift, one would not expect to see any significant phenotypic differences between segregants of different MAC values. Thus, a positive result from the experiments here could resolve both the issue of neutrality for most MAs as called here and the issue of genetic variations or MAC on the complex trait of transgenerational inheritance.

## Results

### Calculation of MAC in the segregants

Using published SNP genotype data for the yeast segragants, we determined the MA frequency or MAF of each SNP in the segregant population of 124 segregants [3-5]. Of the 2956 SNPs scanned, 121 have MAF 0.5 and were considered non-informative. Of the remaining 2835 SNPs, 1589 MAs were from BY and 1246 from RM ($P < 0.01$, Supplementary Table S1). We then calculated the MA content (MAC) in each segregant which ranged from 0.3 to 0.6 (Supplementary Table S1).

We determined whether MAC is simply a measure of the amount of parental alleles in a segregant or whether high MAC values mean more BY alleles since more BY alleles were found as MAs. A segregant could have MAs by either chance or natural selection. If by chance, it would be a high probability event for a BY allele to be present in the high MAC segregants. If by natural selection, it would not be so necessarily. If BY alleles are more deleterious than RM alleles, a segregant of high MAC value may not survive if it is enriched with BY alleles. Indeed, we found that one cannot predict parental allele content from MAC values (Supplementary Table S1). For example, segregant 17_1_a has a low MAC value 0.408 and less BY alleles than RM alleles (1381 vs 1431). On the other hand, another segregant 15_6_c with low MAC value 0.413 has more BY than RM alleles (1584 vs 1245). Segregant 1_1_d has the highest MAC 0.591 and yet has less BY than RM alleles (959 vs 1336, the number does not add to 2835 because some SNPs have no genotype information). Of the 20 segregants with the highest MAC, 9 have less RM alleles than BY alleles while 11 have more RM than BY alleles; and for the 20 segregants with the lowest MAC, 11 have less RM than BY alleles while 9 have more RM than BY alleles (Supplementary Table S1). Such failure of probability theory in predicting the amount of parental alleles from MAC values indicates a non-neutral nature of these alleles.

**Lag phase responses**

If the SNPs or MAs are not neutral as described above, segregants with different MAC would be expected to be different in phenotypic traits, possibly including transgenerational inheritance traits. We tested this by studying the lag phase response trait. There is often a lag phase when microorganisms adapt themselves to new conditions, during which they acquire nutrients from the new growth medium and have strong metabolism level but not yet able to divide [6]. The length of the lag phase is an inherent trait of each organism. We divided the segregant population into two groups of 10 segregants each, the high MAC (HMAC) group with MAC 0.5-0.6 and the low MAC (LMAC) group with MAC 0.3-0.4. The average lag phase of the HMAC group was longer than that of the LMAC group upon acute ethanol treatments (Fig 1, Supplementary Table S2). We then measured the lag phase length after 14 days of adaptive training in ethanol media and 7 days of recovery in normal YPD media. We selected 14 days because time course experiments showed an insignificant change in lag phase after 7 days of training while a similar change to that of 14 days after 30 days of training (data not shown). At the 14th day in ethanol-containing media, the lag phase of all strains became shorter but the decrease was more dramatic for the HMAC group (Fig. 1). The results suggest that HMAC segregants were less able to maintain the inherent trait of lag phase length. The ethanol-trained segregants were next grown in ethanol free YPD medium for 7 days before their lag phase responses to ethanol were measured. All strains showed an increase in lag phase after 7 days of recovery in ethanol free medium (Fig. 1). The HMAC group however showed a more dramatic increase in lag phase length relative to the LMAC group, indicating less stability of the acquired phenotype for the HMAC group.

To show that the acquired phenotype of short lag phase has in fact been memorized for at least certain number of generations, we measured the lag phase length at various time points during recovery from 14 days of ethanol training. We used the two parental strains RM11-1b and BY4716 for this experiment. The acquired phenotype of short lag phase was not lost at 1 or 2 day recovery but disappeared at 8 day recovery (Fig. 2). Yeast typically has a generation time of 2 hours, and so the data indicate that the acquired phenotype of short lag phase can be stably maintained during transgenerational inheritance for at least 24 generations.

To confirm the results from the ethanol treatments, we examined the response of segregants to sodium chloride treatment using the same experimental procedures except that ethanol was replaced by sodium chloride. Overall, the results of the sodium chloride treatment were similar to those of the ethanol experiment, although less dramatic (Fig. 3, Supplementary Table S3).

**MAC and gene expression**

To examine how MAC may affect epigenetic programs, we asked whether MAC could be linked to gene expression patterns by using microarray data from the literature [7]. There were 324 genes in glucose enriched media and 172 genes in ethanol enriched media with significant difference in expression levels between HMAC and LMAC groups at a false discovery rate (FDR) of 10% by SAM analysis (Supplementary Table S4 and S5). As a negative control, three independent random sorting of the 40 segregants did not identify any correlated genes. In glucose enriched media, there were 17 genes expressed higher among the 324 significant genes in the HMAC group (Supplementary Table S4). In ethanol enriched media, 127 genes in the 172 significant genes showed greater expression in HMAC group (Supplementary Table S5). There were 25 genes that were regulated by MAC in both glucose and ethanol conditions (Supplementary Table S6). These observations suggest that MACs are significantly correlated with gene expression profiles and hence epigenetic programs.

**MAC and the number of known additive loci of a trait**

That a complex trait in an individual is correlated with the total amount of SNPs or MAs in the individual suggests a role for multiple genetic loci acting in an additive fashion. To verify this, we took advantage of a published study on a large panel of yeast segregants derived from a cross between a variant strain of BY4716 and a variant of RM11-1a, which identified the number of additive QTLs associated with each studied trait [8]. The number of identified additive loci ranged from 5 to 29 (average 12) for the 46 traits studied, although the study cannot possibly identify all possible QTLs for a trait due to experimental limitations such as sample size. We determined the MAs of the 392 SNPs genotyped for these segregants and calculated the MAC for each of the 1009 segregants (Supplementary Table S7). Of the MAs, 185 were BY alleles and 207 RM alleles ($P > 0.05$). Sixteen traits were significantly linked with MAC by Spearman correlation analysis, and 5 of them remain significant after multivariate regression analysis (Table 1). When the 46 traits were divided into two halves based on their linkage with MAC, the top half with stronger linkage to MAC have on average 14.3 QTLs versus 11.3 for the lower half ($P<0.05$, Student's T test), indicating higher number of additive QTLs for traits linked with MAC (Table 1). Consistently, the 5 traits most definitively linked to MAC as determined by multivariate analysis have on average 16 additive QTLs. The results suggest that the link between quantitative variations of a trait phenotype and MAC is due to the additive effect of a large number of QTLs and the functional difference between MAs and major alleles.

## Discussion

The results here suggest that the effects of environmental factors on inherent traits as well as acquired traits may vary depending on the seemingly normal genetic variations in an organism or cell. Although it is well established that large effect mutations can affect susceptibility to environmental factors, this study may help explain the role of common SNPs or seemingly normal variations in the stability of transgenerational inheritance.

Phenotypic variations in the segregant panel can be partitioned into the contribution of heritable genetic factors and measurement errors or other random environmental effects. In the experiment here, gene–environment interactions should be absent as all the segregants are grown simultaneously under uniform conditions [8].

For the progeny population derived from the cross between BY and RM, the BY alleles were significantly less represented or mostly minor alleles in the 124 strain panel used here. On the other hand, BY alleles were slightly or non-significantly more represented in the 1009 strain panel used in the Bloom et al study [8]. These differences could result from variations in the SNPs ascertained, the parental strains, and the growth conditions, which could all affect the selective pressure on SNPs. That these SNPs are not neutral is supported by both the absence of a correlation between BY allele content and MAC in the 124 segregants panel and the trait difference between HMAC and LMAC groups. They support the previous conclusion on the non-neutrality of most common SNPs [5].

The poor performance of the HMAC group in transgenerational inheritance is in line with the observation that most MAs are minor because they are under slightly negative selection with regard to survival of the segregants under laboratory conditions. It is expected that deleterious MAs would adversely affect some orderly biochemical/physiological pathways. Such notion gives the most parsimonious explanation to the observed result and better predicts segregant properties from their MAC values.

Difference in the stability of lag phase length exists in the parental strains and is hence presumably due to genetic or SNP differences between the strains. However, that difference alone reveals little on whether such difference is due to multiple SNPs and whether SNPs associated with low stability are in general more likely to be deleterious. The results here clarify these issues. MAC linked traits tend to have higher number of identified additive QTLs. While some of these linkages have weak p values, which would be treated as false positives by Bonferroni correction, it is more prudent here to not to use such correction since there is a high risk of false negatives among a few other fallacies [9]. But these weak associations should be verified by future studies, while the study here is more conclusive on the stronger ones as found by multivariate analysis.

If major alleles represent the favored allele to an ordered biochemical pathway, then a mutation that changes a major allele to a minor one could be regarded as a random disruption to the ordered pathway. Thus more minor alleles mean more mutations and hence more randomness or disorder in the system, which may adversely affect certain traits. Most common MAs may not have large effects individually but a group of them together over a threshold limit may have significant effects. The effect could be very minor so that the MAs would not be rare in frequency or under strong negative selection. Most Genome Wide Association Studies (GWAS) or other existing methods may not be able to detect such minor effect SNPs individually and thus create the artificial problem of “missing heritability”[10]. In reality, however, most of what is missing may be in the so called neutral SNPs, whose collective effect can now be detectable by the concept and method of MAC as shown here and elsewhere [5]. Thus, we infer from the effect of MAC that the ethanol induced lag phase response is determined by large number of genetic loci acting in an additive manner. Indeed, ethanol tolerance in yeast is highly heritable and thought to be determined by as many as 251 genes as well as a large number of additive QTLs [11].

Our results further showed a correlation between enrichment of MA contents and mRNA expression, extending previous work on eQTLs [3]. Future work may reveal the mechanisms by which a large number of SNPs or eQTLs may affect the expression of an individual gene.

Most inherent traits and acquired traits are determined by epigenetic programing. There are great variations in the transgenerational epigenetic stability of inherent and acquired traits [1,2]. The work here shows an important role of seemingly normal genetic variations in the stability of transgenerational inheritance. It has implications for disease prevention and treatment. Individuals with more SNP minor alleles may be more susceptible to environmental pathogens due to adverse effect of MAs on inherent traits. But they may also be more easily treatable if treatment was administered relatively early before the disease has progressed past the threshold of no return, because the acquired disease trait may be less stably maintained in these individuals.

**Acknowledgements:**

We thank R. Brem, E. Smith, M. Rockman, and L. Kruglyak for research materials or technical assistance. Supported by the National Natural Science Foundation of China grant 81171880 and the National Basic Research Program of China grant 2011CB51001 (S.H.).

## Materials and methods

**Strains**: The yeast segregants in this study were gifts of Dr. R. Brem and the information on the strains are described in Supplementary Table S1.

**MAF and MAC calculation:** The SNP datasets were obtained from R. Brem, E. Smith, and L. Kruglyak. The MAF of each SNP in the panel of segregants was calculated by PLINK and SNP Tools for Microsoft Excel [12,13]. From such MAF data, we obtained the MA set, which excluded non-informative SNPs with MAF = 0 or 0.5 in the panel. The MA set was equivalent to the genotype of an imagined individual who is homozygous for all the MAs. The MAC of each segregant was then determined by matching the genotype of a sergeant with the MA set; the number of identical genotypes was scored as the number of MAs for the segregant [5]. The MAC of a strain was calculated by dividing the number of MAs carried by the strain by the number of total SNPs scanned.

**Statistical methods**: Gene expression datasets for the segregants were from previous studies [7]. The gene expression difference between the HMAC group and the LMAC group were analyzed using the SAM software. Differences in lag phase length were examined by Student's t test, two tailed. Spearman correlation and multivariate regression analysis were done using GraphPad Prism5 and InStat3.

**Culture conditions:** Segregants were cultivated in the YPD media which consist of 1% yeast extract(OXID), 2% glucose(Sigma), 2% peptone(BD). Ethanol treatment used 7%v/v ethanol in YPD media. Sodium chloride treatment used 0.8M sodium chloride in YPD media. All growth was performed in an Orbital Shaker at 200 rmp and 30℃.

**Growth curve determination:** The segregants were cultivated overnight to OD600=1. Next, every segregant was transferred to 6 ml of fresh medium in a 15 ml round bottom centrifuge tube and was adjusted to OD600 = 0.03 at 200 rmp and 30℃ for 72h (7%ethanol) or 40h (0.8M sodium chloride). The optical density was measured every 2h using Automatic microplatereader at 600nm (OD600). The growth curve was drawn with cultivation time for x axis and ln(OD600) for Y axis. The duration of the period of lag phase can be quantified by the point of intersection that the tangent of the logarithmic phase and the horizon of the original concentration [14]. These segregants were cultivated in the YPD media containing 7%(v/v) ethanol and 0.8M sodium chloride for 14 days (the segregants were transferred to new ethanol media or NaCl media every 48 hours). After 14 days adaptability training, the segregants were cultivated in the normal YPD media for 7 days (they were transferred to the new media every 24 hours and the cells were washed with new media before transfer). Growth curve was then determined at the YPD media containing 7%(v/v) ethanol or 0.8M NaCl. The experiments were repeated three times.

**Figure legends:**

**Figure 1. Lag phase response to ethanol treatment.** The segregants were trained in YPDE (containing 7% v/v ethanol) for 14 days and then recovered in ethanol free YPD for 7 days. The lag phase was measured in YPDE under three conditions (No-training, Training, Recovery training). ** $p<0.001$, * $p<0.05$. Error bars represent standard deviation.

**Figure 2. Epigenetic memory of the acquired trait of short lag phase after ethanol treatment.** Parental yeast strains were used for the experiment and the lag phase length was scored at various time points during recovery from 14 days of ethanol training.

**Figure 3. Lag phase response to sodium chloride treatment.** The segregants were trained in YPDS (0.8M sodium chloride) for 14 days and recovered in YPD for 7 days. The lag phase was measured in YPDS (No-training, Training, Recovery training). * $p<0.05$. Error bars represent standard deviation.

**Table 1. Correlation between MAC and the number of identified additive QTLs of each trait.** The traits were listed in the order of strong to low linkage to MAC as indicated by P values from Spearman correlation.

| Spearman r | P value | P multivariate | Chemicals | # QTLs |
|---|---|---|---|---|
| 0.129 | 0.00001 | | Paraquat | 21 |
| -0.133 | 0.00001 | 0.006 | YNB | 18 |
| 0.109 | 0.0006 | | E6_Berbamine | 15 |
| 0.107 | 0.0009 | | Fluorocytosine | 11 |
| -0.100 | 0.002 | 0.0443 | Tunicamycin | 29 |
| -0.121 | 0.002 | 0.0068 | Raffinose | 7 |
| -0.098 | 0.002 | | Azauracil | 12 |
| 0.094 | 0.003 | 0.0016 | Fluorouracil | 16 |
| -0.088 | 0.005 | | YPD | 17 |
| 0.086 | 0.007 | | Galactose | 9 |
| -0.083 | 0.010 | 0.0015 | Mannose | 10 |
| 0.081 | 0.012 | | Zeocin | 17 |
| -0.079 | 0.013 | | Lactate | 14 |
| -0.073 | 0.021 | | 4NQO | 10 |
| -0.073 | 0.023 | | Magnesium_Chloride | 9 |
| 0.071 | 0.025 | | Cobalt_Chloride | 16 |
| 0.054 | 0.09 | | Lactose | 13 |
| -0.054 | 0.11 | | SDS | 15 |
| 0.046 | 0.15 | | Cisplatin | 13 |
| 0.041 | 0.20 | | Formamide | 10 |
| 0.039 | 0.22 | | Magnesium_Sulfate | 12 |
| 0.032 | 0.32 | | Cycloheximide | 14 |
| 0.030 | 0.35 | | Lithium_Chloride | 22 |
| -0.033 | 0.35 | | Cadmium_Chloride | 6 |
| -0.029 | 0.39 | | YNB:ph8 | 17 |
| 0.030 | 0.40 | | Hydrogen_Peroxide | 6 |
| -0.025 | 0.43 | | Diamide | 20 |
| -0.024 | 0.45 | | Hydroxybenzaldehyde | 10 |
| -0.023 | 0.46 | | Trehalose | 12 |
| 0.023 | 0.48 | | Xylose | 11 |
| -0.028 | 0.49 | | Sorbitol | 8 |
| -0.017 | 0.58 | | YNB:ph3 | 12 |
| 0.016 | 0.61 | | YPD:15C | 10 |
| 0.015 | 0.63 | | YPD:37C | 8 |
| -0.015 | 0.67 | | YPD:4C | 12 |
| -0.014 | 0.68 | | Calcium_Chloride | 13 |

| | | | |
|---|---|---|---|
| 0.013 | 0.70 | Copper | 14 |
| -0.010 | 0.75 | Caffeine | 12 |
| -0.010 | 0.76 | Neomycin | 19 |
| -0.008 | 0.79 | Maltose | 5 |
| -0.001 | 0.97 | Indoleacetic_Acid | 6 |
| -0.001 | 0.97 | Ethanol | 11 |
| 0.001 | 0.98 | Hydroquinone | 10 |
| 0.001 | 0.98 | Menadione | 12 |
| -0.001 | 0.98 | Hydroxyurea | 13 |
| 0.000 | 1.00 | Congo_red | 14 |
| | | P value, top vs bottom half | 0.03 |

**Supplementary Information:**

**Supplementary Table S1.** **List of 124 yeast segregants used.**

**Supplementary Table S2.** **Lag phase length in response to ethanol**

**Supplementary Table S3.** **Lag phase length in response to NaCl**

**Supplementary Table S4.** **Genes regulated by HMAC in glucose enriched media.**

**Supplementary Table S5.** **Genes regulated by HMAC in ethanol enriched media.**

**Supplementary Table S6.** **Genes regulated by HMAC in both glucose enriched medium and ethanol enriched media.**

**Supplementary Table S7.** **MAC and phenotype values of the 1009 segregants panel.**

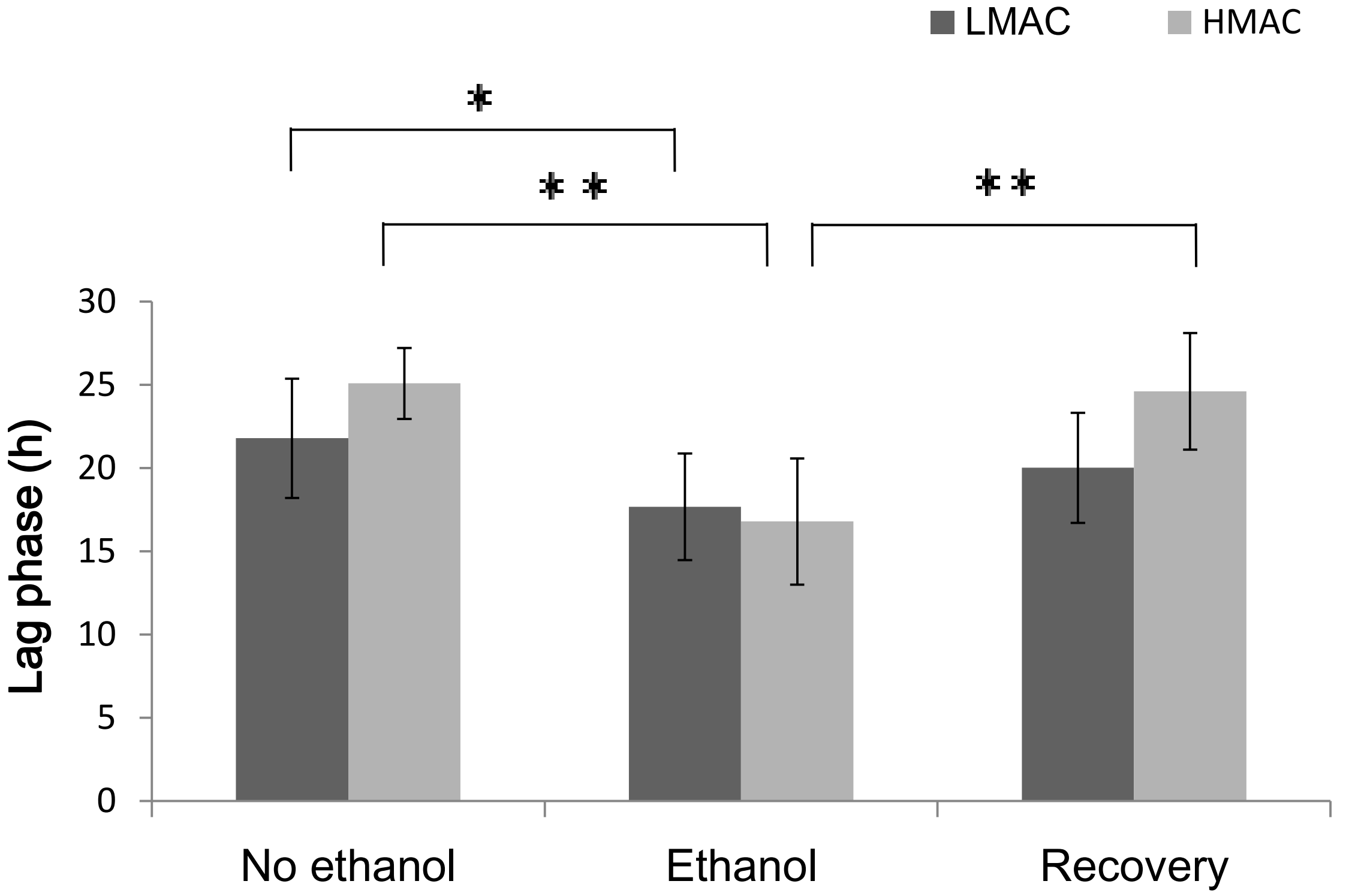

Figure 1

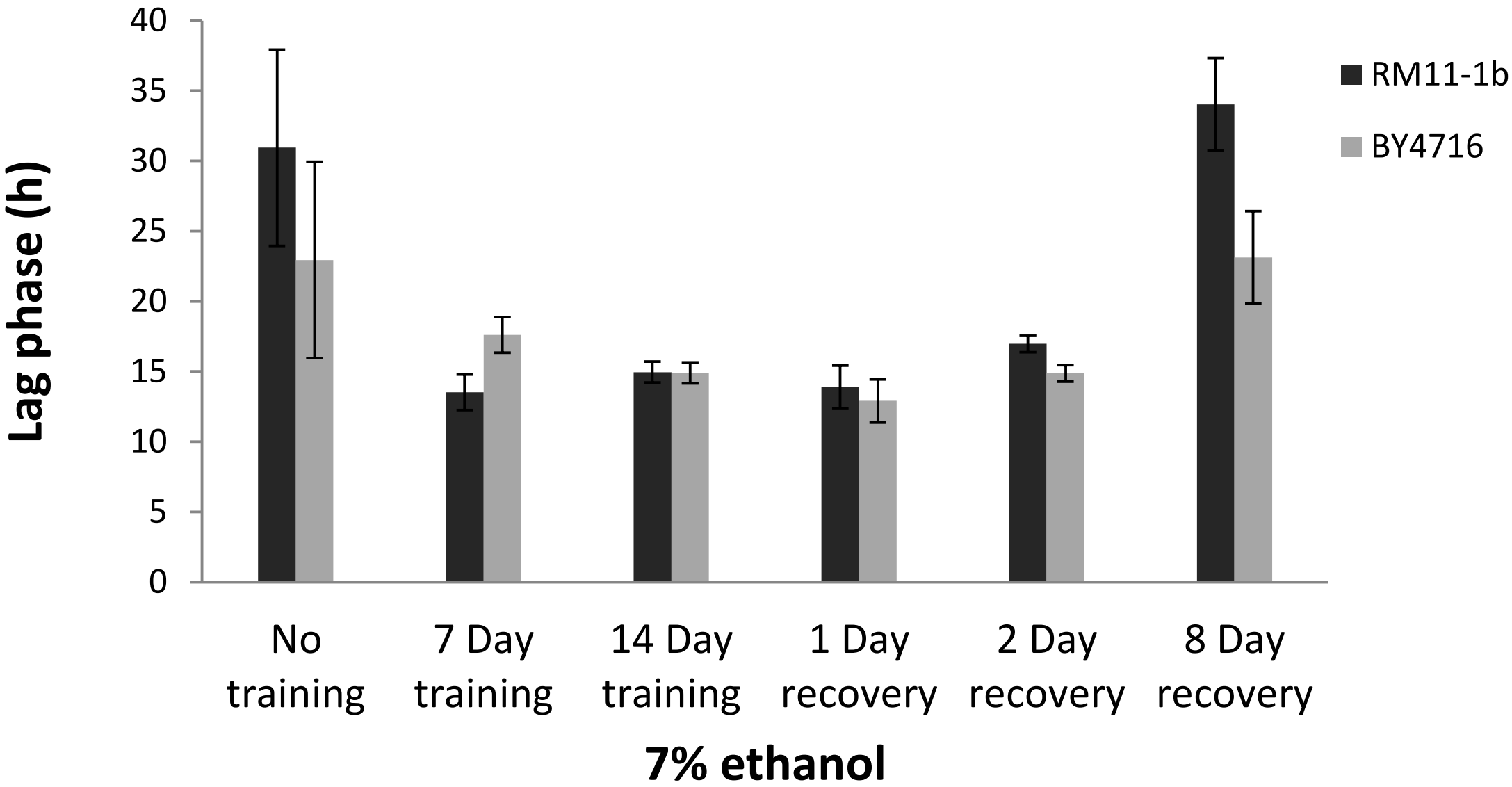

Figure 2

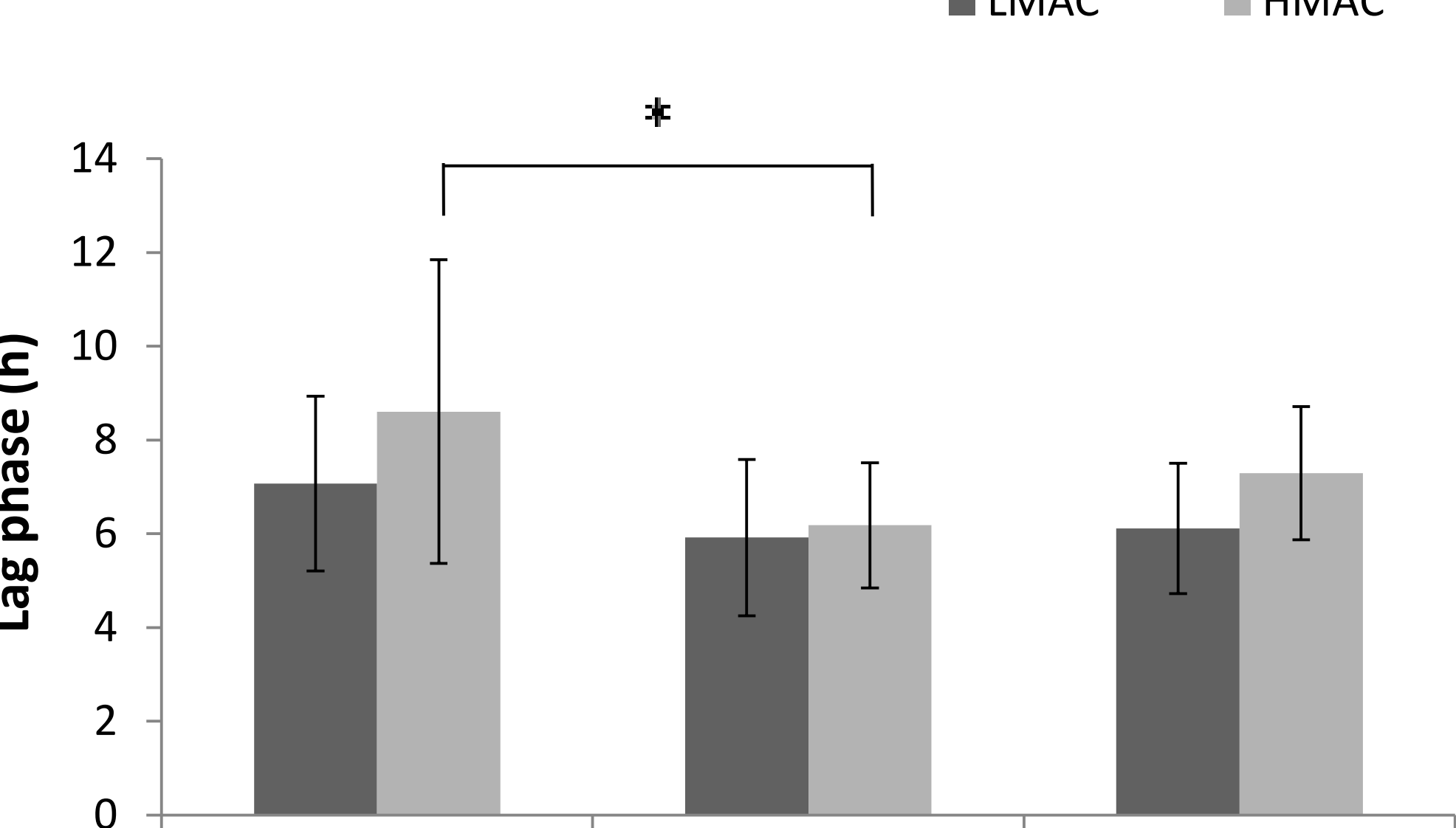

Figure 3

Table 1. Correlation between MAC and the number of identified QTLs of each trait.

| Spearman r | P value | P multivariate | Chemicals | # QTLs |
|---|---|---|---|---|
| 0.129 | 0.00001 | | Paraquat | 21 |
| -0.133 | 0.00001 | 0.006 | YNB | 18 |
| 0.109 | 0.0006 | | E6_Berbamine | 15 |
| 0.107 | 0.0009 | | Fluorocytosine | 11 |
| -0.100 | 0.002 | 0.0443 | Tunicamycin | 29 |
| -0.121 | 0.002 | 0.0068 | Raffinose | 7 |
| -0.098 | 0.002 | | Azauracil | 12 |
| 0.094 | 0.003 | 0.0016 | Fluorouracil | 16 |
| -0.088 | 0.005 | | YPD | 17 |
| 0.086 | 0.007 | | Galactose | 9 |
| -0.083 | 0.010 | 0.0015 | Mannose | 10 |
| 0.081 | 0.012 | | Zeocin | 17 |
| -0.079 | 0.013 | | Lactate | 14 |
| -0.073 | 0.021 | | 4NQO | 10 |
| -0.073 | 0.023 | | Magnesium_Chloride | 9 |
| 0.071 | 0.025 | | Cobalt_Chloride | 16 |
| 0.054 | 0.09 | | Lactose | 13 |
| -0.054 | 0.11 | | SDS | 15 |
| 0.046 | 0.15 | | Cisplatin | 13 |
| 0.041 | 0.20 | | Formamide | 10 |
| 0.039 | 0.22 | | Magnesium_Sulfate | 12 |
| 0.032 | 0.32 | | Cycloheximide | 14 |
| 0.030 | 0.35 | | Lithium_Chloride | 22 |
| -0.033 | 0.35 | | Cadmium_Chloride | 6 |
| -0.029 | 0.39 | | YNB:ph8 | 17 |
| 0.030 | 0.40 | | Hydrogen_Peroxide | 6 |
| -0.025 | 0.43 | | Diamide | 20 |
| -0.024 | 0.45 | | Hydroxybenzaldehyde | 10 |
| -0.023 | 0.46 | | Trehalose | 12 |
| 0.023 | 0.48 | | Xylose | 11 |
| -0.028 | 0.49 | | Sorbitol | 8 |
| -0.017 | 0.58 | | YNB:ph3 | 12 |
| 0.016 | 0.61 | | YPD:15C | 10 |
| 0.015 | 0.63 | | YPD:37C | 8 |
| -0.015 | 0.67 | | YPD:4C | 12 |
| -0.014 | 0.68 | | Calcium_Chloride | 13 |
| 0.013 | 0.70 | | Copper | 14 |
| -0.010 | 0.75 | | Caffeine | 12 |
| -0.010 | 0.76 | | Neomycin | 19 |
| -0.008 | 0.79 | | Maltose | 5 |
| -0.001 | 0.97 | | Indoleacetic_Acid | 6 |
| -0.001 | 0.97 | | Ethanol | 11 |
| 0.001 | 0.98 | | Hydroquinone | 10 |
| 0.001 | 0.98 | | Menadione | 12 |
| -0.001 | 0.98 | | Hydroxyurea | 13 |
| 0.000 | 1.00 | | Congo_red | 14 |
| | | P value, top vs bottom half | | 0.03 |